 \documentstyle[11pt,aaspp4]{article}  % for astro-ph, change \end{document}
\slugcomment{ApJ Lett, submitted, Jan 31, 1999}

\lefthead{Fenimore et al.}
\righthead{GRB990123: External or Central Engine?}
\begin{document}
\title{GRB990123: Evidence that the Gamma Rays
Come from a Central Engine}
%see sample2.tex for adding allification footnotes
\author{E.~E.~Fenimore$^{1}$, E.~Ramirez-Ruiz$^{1,2}$, and Bobing 
Wu$^{1,3}$}  % tbd: affil
\affil{$^1$MS D436, Los Alamos National Laboratory, Los Alamos, NM 87545}
\affil{$^2$Facultad de Ciencias, Universidad Nacional Aut\'onoma de
M\'exico, Distrito Federal, M\'exico 04510}
\affil{$^3$Institute of High Energy Physics, CAS, Beijing 100039, China}
%%%  QQ  %%
\begin{abstract}
GRB990123 was a long complex gamma-ray burst with an optical transient
that started early within the gamma-ray phase (\cite{carlgcn,carl}). The
peak and power law 
decay of the early optical emission strongly indicates the presence of a 
decelerating relativistic shell during that phase.
Prior to this burst, it was not known if the shell decelerated during the 
burst, so an external shock origin for the gamma rays was still possible. If 
the gamma-rays are produced in the external shock, then the pulse widths 
should reflect the observed deceleration of the shell and 
increase by about  2.3.  We analyze the fine time structure 
observed in the gamma-ray data from  BATSE  and determine that the 
width of the peaks do  not increase as  expected for a decelerating shell; 
the later pulses are, at most, a factor of 1.15 longer than the earlier 
pulses.
We also analyze the variability to determine what fraction of the 
shell's surface could be involved in the production of the gamma 
rays, the so-called surface filling factor. 
For GRB990123 we find a filling factor of 0.008.
The lack of pulse width evolution eliminates the only remaining 
kinematically acceptable external shock explanation for the gamma-ray 
phase and, thus, the gamma rays must originate at a central engine.
\end{abstract}


\newcount\eqnumber
\eqnumber=1
\def\neweq{{\the\eqnumber}\global\advance\eqnumber by 1}
\def\eqnam#1#2{\xdef#1{\the\eqnumber}}
\def\lasteq{\advance\eqnumber by -1 {\the\eqnumber}\advance
    \eqnumber by 1}
\def\domega{{\rm d}$\Omega$}
\def\Mesz{M\'esz\'aros}
\def\INITIALCON{{4\pi E_{54} \over \rho_0{\rm d}\Omega}}
\section{INTRODUCTION}

On January 23, 1999, the Robotic Optical Transient Search Experiment
(ROTSE) discovered strong optical emission (9th mag) during a gamma-ray
burst (\cite{carlgcn,carl}). Remarkably, such extraordinary behavior was
predicted a few weeks before (\cite{sp99a}).  This event, GRB990123, had
a location from the BeppboSAX satellite (\cite{pirogcn,heiseiau}) and a
measured redshift of the optical transient of $z = 1.6$
(\cite{kelsoniauc,hjorthgcn}).  The Burst and Transient Source Experiment
(BATSE) observed the burst (\cite{kippengcn}), as did the Comptel
experiment (\cite{connorsgcn}). 

The detection in Comptel implies that the burst had a typical hard
gamma-ray burst (GRB) spectrum.  GRB spectra often extend to very high
energies with no indication of attenuation by photon-photon interactions.
This implies substantial relativistic bulk motion of the radiating
material with Lorentz factors in the range of $10^2$ to $10^3$. Two
classes of models have arisen that explain various aspects of the
observations. In the ``external'' shock models (\cite{mr93}), the release
of energy is very quick, and a relativistic shell forms that expands
outward for a long period of time ($10^5$ to $10^7$ s). At some point,
interactions with the external medium (hence the name) cause the energy of
the bulk motion to be converted to gamma-rays. Although the shell might
produce gamma-rays for a long period of time, the shell keeps up with the
photons such that they arrive at a detector over a relatively short period
of time. If the shell has a velocity, $v = \beta c$, with a corresponding
bulk Lorentz factor, $\Gamma = (1-\beta^2)^{-1/2}$, then photons emitted
over a period $t$ arrive at a detector over a much shorter period, $T =
(1-\beta)t = t/(2\Gamma^2)$. Although this model is consistent with the
short energy release time expected for a compact object merger and the
observed long time scale of GRBs, we have argued that it cannot explain
gamma-ray emission with gaps. It can only explain the rapid time
variability if the shell is very narrow, has angular structure much
smaller than $\Gamma^{-1}$, and cannot be decelerating
(\cite{fmn96,enrico}).

The alternative theory is that a central site releases energy in the form
of a wind or multiple shells over a period of time commensurate with the
observed duration of the GRB (\cite{rm94}). The gamma-rays are produced by
the internal interactions within the wind; hence these scenarios are often
referred to as internal shock models although they might actually involve
the Blandford-Znajek effect (\cite{mr97,pac97}). The discovery of x-ray
afterglows lasting hours (\cite{cos97}), optical afterglows lasting weeks
to months (\cite{metzger97}), and radio afterglows lasting many months
(\cite{frail97}) strongly indicated external shocks were involved in some
aspects of GRB emission. The observed power law decay of the afterglows is
expected from many external shock models
(\cite{wrm97,danr97,tavani97,wkf98,mrw98,rm98,spn98}). 

\cite{ps97n} suggested that the initial gamma-ray phase is due to internal
shocks from a relativistic wind (or multiple shells) that merge into a
single relativistic shell which then produces the afterglows in a manner
similar to the external shock models. More recently, \cite{sp99an} have
predicted bright early optical afterglows resulting from a reverse shock
in the shell. Their detailed calculations showed behavior much like that
observed in GRB990123: a rapid rise in the optical soon after the burst
starts to very bright levels and a power law decay as the shell
decelerates. 

Using kinematic arguments, the rapid time variability of many GRBs
indicates that only a small
fraction (typically 0.005) of the shell surface can be involved in the
gamma-ray emission (\cite{fmn96,sp97,fcrs99}). There
might be ways to obtain
filling factors as large as 0.1 (\cite{dermer99}). However, the average
pulse width is remarkably constant (to within a few percent) and, thus,
shows no sign of any deceleration (\cite{enrico}).  As a result, the 
only kinematically allowed
external shock models for the gamma-ray phase have been forced to involve
very narrow shells and no deceleration.

The early optical emission of GRB990123 was discovered by ROTSE
(\cite{carlgcn,carl})  which consists of four 11.1 cm aperture telephoto
lenses with unfiltered CCD cameras on a single rapidly slewing mount. It
is capable of responding rapidly to alerts from the Gamma-ray Coordinates
Network (GCN). In the case of GRB990123, the initial trigger was sent out
on GCN well before the explosive rise in the gamma rays. Optical
exposures were actually taken while the gamma-ray emission was still
increasing.  The ROTSE experiment detected extremely strong optical
emission that increased rapidly from 12th to 9th magnitude, then faded
with a power law decay during the burst (\cite{carl}). Other optical
observations many hours later saw a continued power law decline beyond 
the initial ROTSE data (\cite{bloomgcn206}).

 Figure 1 shows
the ROTSE optical observations overlaid on the BATSE time history of the
gamma-ray emission. Both are plotted as Log-Log, an unusual way to present
gamma-ray time histories.  This presentation shows the power law decay of
the optical data, and it allows one to see how the gamma-ray data tracks
the optical emission. The gamma-ray envelope would also show a power law
decay if, it too, were due to a relativistic shell.  Interestingly, the
gamma-ray emission does have an envelope that is similar to the optical
envelope.  However, one must be careful because the ROTSE observations
were short compared to their separation, so only a few points within the
time history are sampled.  In fact, the optical emission could have
closely mirrored the three main gamma-ray releases of energy (at 25,
37, 80 s) and ROTSE
would not have resolved them.
The potential similarity
of the
optical and gamma-ray envelopes means one cannot necessarily  argue
from the envelopes alone for a
different origin. 

 The purpose of this paper is to analyze the BATSE data to show that the
fine time structure of the gamma-ray emission does not change during the
burst as it should if the gamma rays originate on the decelerating shell
as indicated by the optical emission.
We find  there is effectively no
apparent deceleration in the gamma-ray source.
Thus, we
eliminate the only remaining external shock model for the gamma-rays and
conclude that the gamma-rays in this source come from a central engine
while the optical emission arises from decelerating external shocks.

\section{VARIATION IN PULSE WIDTH IN GRB990123}

In the early phase of the shell's expansion, $\Gamma$ is effectively
constant ($=\Gamma_0$).  Eventually, the shell begins to decelerate as it
sweeps up the interstellar medium (ISM). Without detailed knowledge of the
physical process that generates the gamma rays, it is not clear if the
expansion is adiabatic or dominated by radiation losses.  However, both
solutions lead to a power law decay of the Lorentz factor
(\cite{spn98,mrw98,rm98}). For typical parameters, the decay is $\Gamma(T) 
\propto T^{-3/8}$ although other indexes are possible (see
\cite{spn98,mrw98,rm98}).  \cite{sp99an} predicted that the
very 
early deceleration could be somewhat slower: $\Gamma(T) \propto T^{-1/4}$ 
in a 
long plateau before the $\Gamma^{-3/8}$ phase, although this depends on 
the initial conditions and  was not observed in this burst.

The proper time for a gamma-ray pulse in our rest frame, $\Delta t$,
should be measured by
clocks placed in our rest frame at all emitting sites, clearly, an
impossible task. Rather, we have one clock (i.e, the BATSE detector) that
measures when the photons arrive at a single location in our rest frame. 
We
denote the arrival time as $T$. 
Rather than a Lorentz transformation, these two ``times'' are related by
$\Delta T = (1-\beta\cos\theta)\Delta t$ where $\theta$ is the
angle between our line of sight and the region on the shell which is
emitting. The relationship between time measured by a clock moving with
the shell ($\Delta t'$) and our rest frame time {\it is} a Lorentz
transformation: $\Delta t' = \Gamma^{-1}\Delta t$.  Thus, BATSE time is
related to time measured in the rest frame of the shell as
\eqnam{\TT}{TT}
$$
\Delta T = \Gamma(1-\beta\cos\theta)\Delta t'~~.
\eqno(\neweq)
$$
Here, we have ignored
cosmological terms  because they do not introduce any differential
effects.

 The average profile of GRBs displays a fast rise and a slower  
decay. This profile is often abbreviated as a ``FRED''
(fast
rise, exponential decay) although the actual average decay is linear
(\cite{fen99}). The fast rise
indicates that the shell emits only
over a small range of radii and one is observing emission from near the 
line of sight early in the burst. The slow decay
is due to the late arrival of emission from regions off the line of sight.
Using Equation (\TT) evaluated at $\theta =0$ and $\theta = \Gamma^{-1}$,
the later emission from a shell ought to have pulses that are about
twice as long as the pulses at the beginning of the burst.  If, in
addition, the shell is slowing down, the $\Gamma$ dependency would cause a
commensurate increase in the pulse width.  An analysis of 53 bright BATSE
bursts showed that the pulse width is remarkably constant (a few percent) 
throughout the gamma-ray emitting phase of a GRB. That
constancy strongly indicates that the only kinematically acceptable
external shock model for the gamma-ray burst phase is one that has no
deceleration and comes from a range of angles that is much smaller than
$\Gamma^{-1}$ (\cite{enrico}).

The GRB990123 observations eliminate that last remaining external shock
model.  In this burst, we see an optical signal that peaks and decays during the
gamma-ray emitting phase.
Either would be a clear indication that the shell has started to
deacelerate.
 In our previous work, we showed
the lack of time variability two ways.  First, we showed that the average
aligned peaks of many bursts is virtually identical throughout the
$T_{90}$ period. (Here, $T_{90}$ is the duration containing 90\% of the
gamma-ray photons.) Second, we used the fits to pulses by \cite{norris96n}
to show that there is no trend in individual bursts to have peaks wider late
in the gamma-ray phase.  Since GRB990123 is an individual burst, it would
be best if the individual peaks were fit until all variations have been
accounted for and then determine if a trend is present.  However, the
complexity of the overlap between peaks in GRB990123 would probably
prevent fitting every peak (see \cite{norris96}). 

To determine the variability of the time scale in GRB990123, we have
analyzed four regions labeled A - D in Figure 1.  In each region, we
first removed the envelope of emission by subtracting the time history
smoothed by a boxcar function (width = 2 sec).  An autocorrelation of the
residuals (see Fig. 2) has a width that is related to the average time
structure in each region. If the pulses were, on average, wider later in
the burst, one would observe  autocorrelation functions for A - D that
progressively get wider.  Rather, sometimes A is wider, sometimes D is
wider.  The maximum spread of $\Delta T_{\rm D}/\Delta T_{\rm A}$ when  the
autocorrelation is greater than 0.5, is only 1.15.  However, when the
autocorrelation function is 0.25, $\Delta T_{\rm D}$ is actually narrower than
$\Delta T_{\rm A}$, by a factor of $\sim 1/1.15$ (and $\Delta T_C$
and $\Delta T_B$ are even narrower than $\Delta T_{\rm A}$).  On average,
$\Delta T_{\rm D}/\Delta T_{\rm A}$ is about unity
and we are confident that its
value is less than the maximum
observed value of 1.15.

  There are two sources of pulse width from an external shell: angular 
effects and deceleration.  The pulse width at two times A and D are 
related as
\eqnam{\WIDRELATE}{WIDRELATE}
$$
{\Delta T_{\rm D} \over \Delta T_{\rm A}} =
{\Gamma_{\rm D} (1-\beta\cos\theta_{\rm D}) \over
\Gamma_{\rm A} (1-\beta\cos\theta_{\rm A})}
\eqno(\neweq)
$$
where $\theta_{\rm D}$ and $\theta_{\rm A}$ are the angles
responsible for the
emission.
If the emission is from a relativistic shell that turns on for a short 
period after expanding away from the central site, then (from Eq. [5] of 
\cite{fmn96}):
\eqnam{\TLAM}{TLAM}
$$
\Gamma(1-\beta\cos\theta_{\rm D}) = {T_{\rm D} \over 2\Gamma T_0}
\eqno(\neweq)
$$
where $T_{\rm D}$ must be measured from when the shell left the central site
and $T_0$ is the time of the peak of the emission.  
During the afterglow, $\Gamma(T) \propto T^{-3/8}$ so
\eqnam{\PULRAT}{PULRAT}
$$
{\Delta T_{\rm D} \over \Delta T_{\rm A}} =
\bigg[{T_{\rm D} \over T_{\rm A}}\bigg]^{11/8}
~~.\eqno(\neweq)
$$
It is likely that the shell started to leave the central site at about 
the time when the first gamma rays were emitted.  BATSE detected emission 
quite early in this burst, so we will use the BATSE time for $T$. Time
period A is at $\sim 45$ s after the start and period D is at $\sim 82$ s. 
Based 
on this, we expect the pulse widths to increase by about a factor of 2.3,
much larger than observed, and very easy to detect if it was present.
\cite{dermer99n} did detailed simulations of GRB pulses produced by
external shocks on a decelerating shell. These simulations (e.g., Fig. 2
in \cite{dermer99}) show pulses that are, indeed, more than a factor of 2
wider late in the burst.  This is not observed in most GRBs and not in
GRB990123.

\section{SURFACE FILLING FACTOR FOR GRB990123}

Another argument against a shell model is given
by an analysis of the ``surface
filling factor''.
We define the surface filling factor as the fraction of the
shell's surface
that becomes active.  Let $A_N$ be the area of an emitting entity and 
$N_N$ be
the number of entities that (randomly) become active during an 
observation period,
$T_{\rm obs}$.  If $A_{\rm obs}$ is the area
of the shell that can contribute
during $T_{\rm obs}$, then the surface filling factor is
\eqnam{\EFF}{EFF}
$$
f = N_N{A_N \over A_{\rm obs}}~= ~~N_N{A_N \over \eta A_S}~
\eqno(\neweq)
$$
where $\eta$ is the fraction of the visible area of the shell ($A_S$) that
contributes during the interval $T_{\rm obs}$ (approximately 1).
The rapid variations in GRB time histories imply emitting entities the
size of $\Delta R_\perp \sim c\Gamma\Delta T_p$.  Assuming a single
 expanding shell, these entities must form on a much larger
surface, $\sim c\Gamma T$.

 There are three  
cases. Case a is the constant $\Gamma$ phase,
case b is the initial deceleration when the size of 
the shell ($\rm{ d}\Omega$) exceeds the radiation beaming angle, and case 
c is when the deceleration reduces the beaming such 
that  the shell's angular size  is no longer larger 
than the beaming angle.
For these cases, the filling factor is related to the observations by
\eqnam{\EFFEQ}{EFFEQ} $$ 
f = \cases{ N_N \big[{\Delta T_p \over T}\big]^2 {1 \over k\eta}
& case a: $\Gamma(T) = \Gamma_0$,  \cr
N_N {\rm d}\Omega^{-1}~
10^{-6}~
\big[\INITIALCON\big]^{-1/4}
{\Delta T_p^2 \over k\eta T^{5/4}}
& case b: ${\rm d}\Omega > 2\pi(1-\cos\Gamma^{-1})$,  \cr
N_N ~{0.07 \over k\eta}~\big[{\Delta T_p \over T}\big]^2
& case c: ${\rm d}\Omega < 2\pi(1-\cos\Gamma^{-1})$,  \cr
} \eqno(\neweq)
$$
where $k$ is 16 if the pulses are the result of the shell interacting with 
ambient objects (e.g., clouds) and 13 for entities that undergo
causally connected growth (e.g. shocks, see \cite{fcrs99n} for
details). Here, the term $\INITIALCON$ 
relates the characteristics of the shell to how fast it slows down. 
$E_{54}$ is the kinetic energy of the shell in units of $10^{54}$ erg and 
$\rho_0$ is the density of the material the shell runs into.

Basically, $N_N$ is the number of individual peaks in the time history
but, because peaks often overlap, one cannot just count the number of
peaks. Rather, the number of entities can be estimated from the
fluctuations in the time history under the assumption that the
(non-counting statistics) fluctuations are due to a randomly varying number
of underlying entities.  In any random process, the square of the mean
divided by the root-mean-square is approximately the rate of occurrence,
$\mu_N$.  We remove the envelope of emission (by either fitting a function
or smoothing the time history), and determine $\mu_N$ from the
fluctuations
in the residuals.  Then, $N_N$ is $\mu_N \Delta T_p/T$, where $\Delta T_p$
is the typical pulse width for which we use 0.3 s (from Fig. 2).

In Figure 3 we show the distribution of surface filling factors
as a function of
burst duration $T_{50}$ based on BATSE bursts. The solid squares are
FRED-like bursts, and
the open squares are long complex bursts.  Although some of the smooth
FRED-like bursts can have surface filling factors near unity,
most bursts have
values on the order of $5 \times 10^{-3}$.

The solid circle is the filling factor for GRB990123.  We have used 
equation (\EFFEQ a) even though the optical emission indicates that the 
shell is decelerating. Equations (\EFFEQ b and c) are for decelerating 
shells and would give even smaller 
values. GRB990123 has a very typical value for the filling 
factor (0.008) implying that there are many fewer emitting entities than the 
minimum number of possible emitting sites.  This forces one to conclude 
that the gamma-ray emitting shell must have angular structure much smaller 
than $\Gamma^{-1}$ and that only a portion of the shell becomes 
gamma-ray active.  It does not, however, necessarily 
indicate that more energy is needed in the reservoir (\cite{fcrs99}).

\section{DISCUSSION}

Although 9th magnitude optical emission seems incredible from an object at
$z \sim 1.6$, it was predicted a few weeks before the event
(\cite{sp99a}).  The low time resolution of the ROTSE data prevents
detailed comparisons to the theory, but the agreement is remarkable. 
\cite{sp99an} predicted early optical emission of up to 7th magnitude,
and ROTSE saw at least 8.95 in this burst. They also predicted a fast rise
with a power law slope of up to 3.7, and the first two ROTSE points can be
connected by a power law with a 3.5 index.  The ROTSE optical peak is
at 45 s from the start of the event and \cite{sp99an} predicted 30 to
50 seconds, depending on the initial Lorentz factor.  The agreement is not
perfect, however. They predict peak times of 30 - 50 s for short bursts 
whereas this is a long burst.  Much better agremment is 
achievable with the proper selection of parameters (\cite{sp99b}).
Nevertheless, we feel that the points of agreement that exist with the
theory are additional evidence that there is a decelerating external shock
during the gamma-ray phase. 

  If the gamma rays are produced with an external shock, the resulting
pulse width should increase.  Analyses of previous bursts showed no such 
trend forcing one to
conclude that the only viable external shock model had to involve very
narrow shells and no deceleration (\cite{enrico}). GRB990123 appears to be
a normal GRB, with normal time variations as witnessed by a normal filling
factor (see Fig. 3). From
the optical and gamma-ray data, we estimate that the pulse width should
increase by about a factor of 2.3 from a combination of angular and
deceleration effects. Averages from many bursts show an increase of only a
few percent (\cite{enrico}). From this single burst, we find changes
less than 15\% (see Fig. 2).  We conclude that the gamma-rays are not
coming from the decelerating shell but from the central site.

We have shown that the source region for the gamma rays must be small to
explain the variability. Since our limitation is based on kinematics, we
cannot comment on the physical process that generates that gamma-ray
phase. In particular, we cannot say that internal shocks are the origin of
the gamma rays.  However, internal shocks from a wind seem to require
large variations in the $\Gamma$ of the shells that collide ($\sim$ a
factor of 2).  It is surprising that such a large variation produces no
effects on the resulting pulse widths.  Thus, we conclude that the source
is small (a ``central engine'') but not that it is necessarily
powered by internal
shocks.

\acknowledgments We thank Galen Gisler and Hui Li for useful comments on
this manuscript. Gerry Fishman and the BATSE team provided the GRB990123
time history.  This work was done under the auspices of the US Department
of Energy.

%
% Option 1.  Using this option, only the figure captions are included in the
% main body of the manuscript.  The figure captions must start on a new page.
% The captions are generated with the \figcaption[]{} command: the first 
% argument is optional, if you put something in there, put the name of the 
% EPS file that goes with the caption; the second argument is the figure 
% caption itself, and may include a \label command.  The \figcaption command
% generates the figure numbers.  This option is acceptable for all manuscript
% submissions.

\clearpage
%
% figure 3 is based on the figure is eff_asca_apj.tex and used values made
%  by [.batse]reformat_enrico_eff.for and mongo-ized by
%  mongo_eff_asca_fig3.inst.  This version was made by
%  [.batse]mongo_grb990123_fig3.inst
%
\figcaption[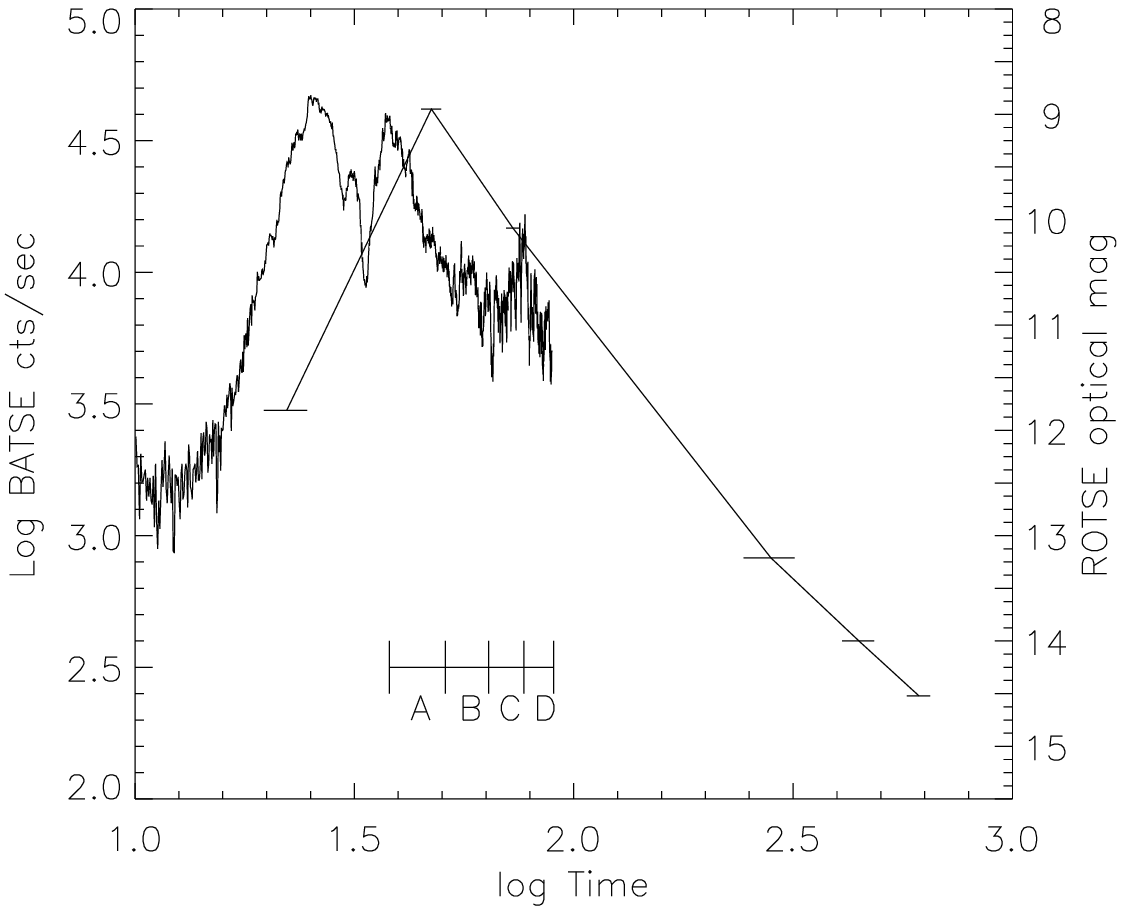]{
The BATSE and ROTSE time history for GRB990123 as a function of Log time 
from the BATSE trigger. The ROTSE data is from Akerlof, et al.~(1996a)
and the BATSE
data is courtesy of Gerry Fishman. The right axis is for the optical 
emission and the left axis is 
for the gamma-ray intensity. Both are logarithmic to show the power law 
decay in the optical, and so one can compare the emission of the 
gamma-rays to the optical.
 Although distorted by the random occurrence of
peaks,  the BATSE emission could have a decay similar to the optical 
emission.  The horizontal bars on the ROTSE samples indicate the duration 
of each observation.  Thus, the ROTSE data is sparse and one should not 
assume that the optical emission follows the lines connecting the ROTSE 
points which are there only to guide the eye.  The average time 
structure during the horizontal periods labeled A - D do not show the 
variation that one would expect from a decelerating shell.
 \label{fig1}}

\figcaption[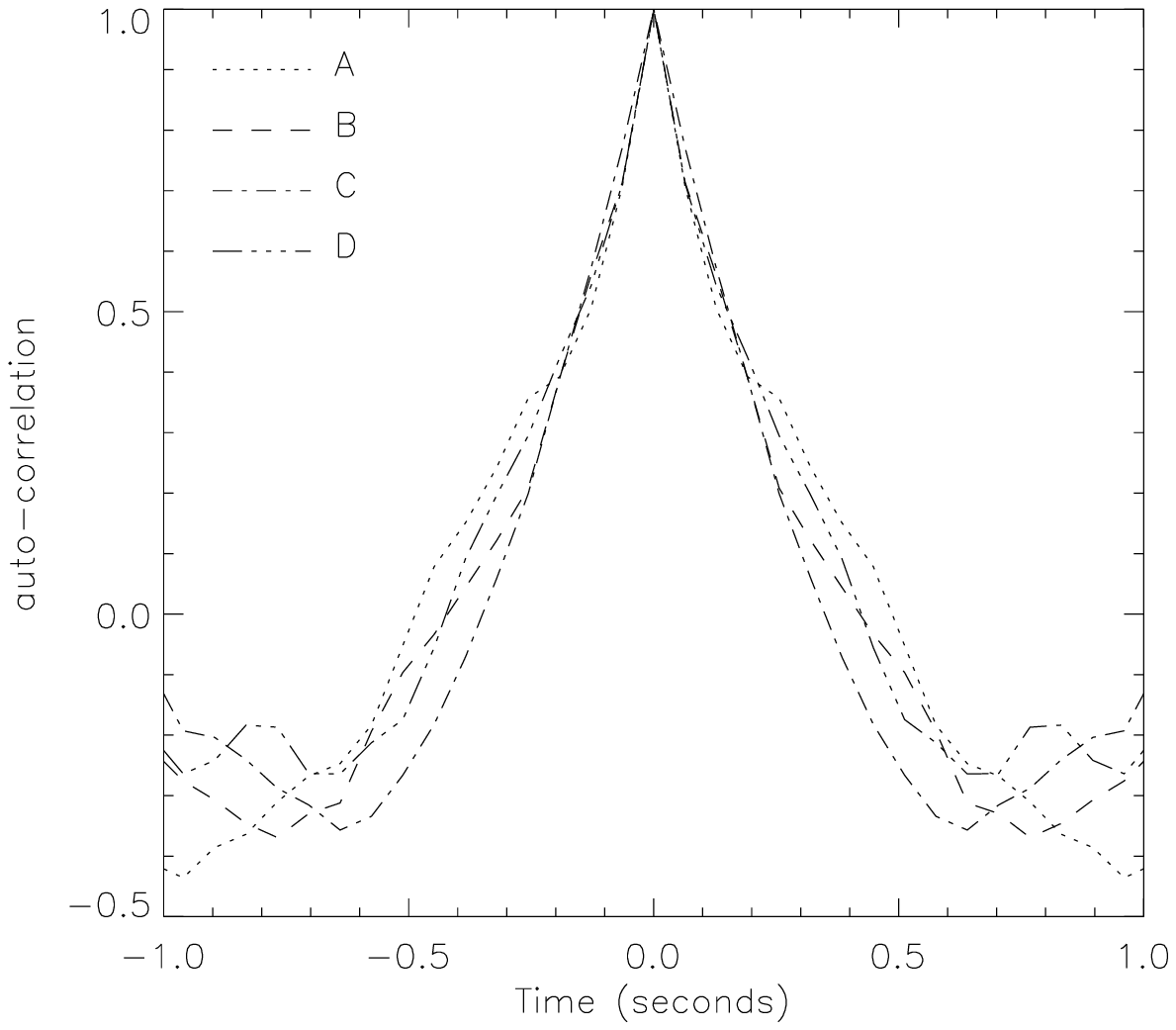]{
The autocorrelation of the  BATSE time history (with the envelope 
removed) of GRB990123 for periods A - D 
in Figure 1.  If the gamma-rays originated on a decelerating shell 
extending over angles  $\sim\Gamma^{-1}$, then one would expect 
that the later pulses (period D) would be wider than the earlier pulses 
(period A) by about 2.3.  This is not observed, the maximum spread is
about 1.15.
\label{fig2}}

\figcaption[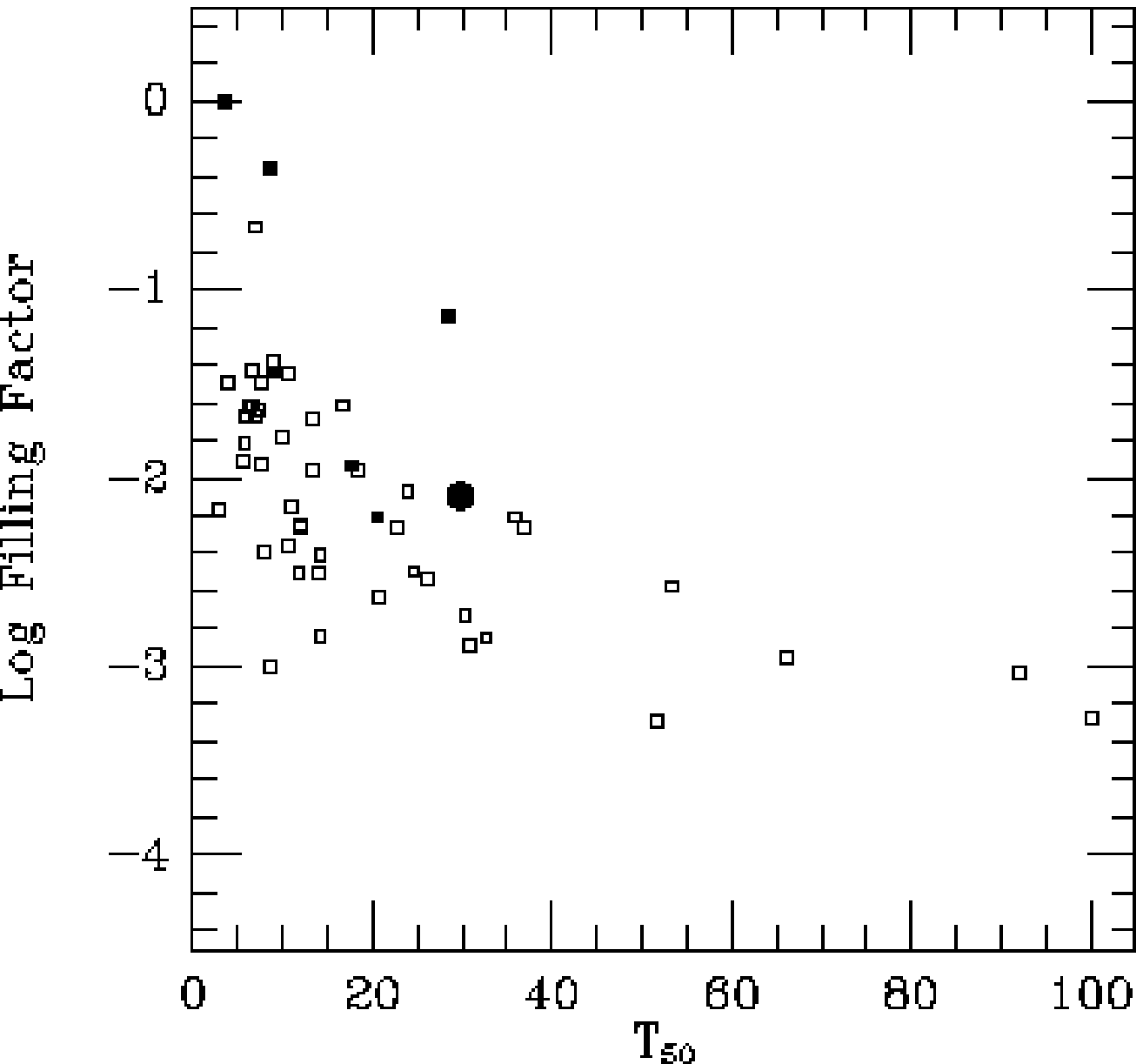]{ Typical values of the fraction of a
relativistic shell that become active during a GRB as a function of the
duration of the emission ($T_{50}$) if the gamma-ray come from an 
external shock.  The six solid squares are FRED-like BATSE bursts 
and the 46 open squares are long, complex BATSE bursts
(from Fenimore, et al.~1999).
The large solid circle is for GRB990123 found from Equation (\EFFEQ a). 
\label{fig3}}

% \end{document}

% Option 2.  The figure captions are printed on a caption page(s) as in 
% option 1.  The figures available as EPS files are then printed at the
% end of the document, one figure per page, using the \plotone command.
% If you wish to process this option then simply comment out the \end{document}
% just above these five lines. 

\clearpage

\centerline{Figure 1}
\plotone{op_batse.eps}
\clearpage

\centerline{Figure 2}
\plotone{grb990123_fig2.eps}
\clearpage

\centerline{Figure 3}
\plotone{grb990123_fig3.eps}
\clearpage

\end{document}